\documentclass[oneside,11pt]{article}

\usepackage{graphicx}
\usepackage{dcolumn}
\usepackage{bm}
\usepackage{multirow}
\usepackage{fullpage}
\usepackage[utf8]{inputenc}
\usepackage[T1]{fontenc}
\usepackage{mathptmx}
\usepackage{etoolbox}
\usepackage{physics}
\usepackage{amssymb,amsmath}
\usepackage{siunitx}
\usepackage{mathtools}
\usepackage{titlesec}
\usepackage{titling}
\usepackage{upgreek}
\usepackage[backend=bibtex,sorting=none,style=phys]{biblatex}
\bibliography{main}

\usepackage{xr}

\makeatletter
\def\@email#1#2{
 \endgroup
 \patchcmd{\titleblock@produce}
 {\frontmatter@RRAPformat}
 {\frontmatter@RRAPformat{\produce@RRAP{*#1\href{mailto:#2}{#2}}}\frontmatter@RRAPformat}
 {}{}
}
\makeatother
\begin{document}
\titlespacing\section{5pt}{12pt plus 4pt minus 2pt}{5pt plus 2pt minus 2pt}
\titlespacing\subsection{5pt}{12pt plus 4pt minus 2pt}{5pt plus 2pt minus 2pt}
\titlespacing\subsubsection{5pt}{12pt plus 4pt minus 2pt}{5pt plus 2pt minus 2pt}

\title{Coupled-cluster approach to Coster-Kronig decay and Auger decay in hydrogen sulfide and argon}
\author{Jan Philipp Drennhaus, Anthuan Ferino-Pérez,\\ Florian Matz, and Thomas-C. Jagau\\E-Mail: \texttt{thomas.jagau@kuleuven.be}\\Department of Chemistry, KU Leuven, B-3001 Leuven, Belgium}

\date{\today}

\maketitle

\begin{abstract}
We perform \textit{ab initio} simulations of the total and 
partial Auger decay widths of 1s$^{-1}$, 2s$^{-1}$ and 2p$^{-1}$ 
ionized hydrogen sulfide and 2s$^{-1}$ ionized argon with non-
Hermitian quantum chemistry. We use coupled cluster theory with 
single and double substitutions (CCSD) and equation of motion 
CCSD (EOM-CCSD) and discuss the novel application of (equation 
of motion-) second order M\o ller-Plesset perturbation theory 
(MP2). We find good agreement between the methods for the 
1s$^{-1}$ hole of H$_2$S, whereas for the other holes we can 
only use the EOM methods. We obtain very large decay widths of 
the 2s$^{-1}$-vacant states due to intense Coster-Kronig 
transitions with excellent agreement to experiments. The three 
2p$^{-1}$ holes show completely different spectra because a 
decay channel is only significant when one of the final holes is 
spatially aligned with the initial hole. Lastly, we observe that 
triplet channels are much more important for the 2s$^{-1}$ and 
2p$^{-1}$ holes than for the 1s$^{-1}$ hole, for which it is 
well known that triplet channels only contribute weakly to the 
total Auger intensity. 
\end{abstract}

\section{Introduction}
The Auger-Meitner effect\cite{meitner22,auger23} is the dominant 
relaxation mechanism of core-vacant states of light elements, i.e., nuclei lighter than about germanium for K-shell holes and 
lighter than about neptunium for L-shell holes. This is 
equivalent to an ionization energy range of up to 13 
keV\cite{bambynek72,thompson09,hitchcock94}. Auger decay is an 
autoionization of atoms or molecules via emission of Auger 
electrons, driven by the simultaneous filling of the core-hole. 
Core-vacant states are not only byproducts of irradiation with X-
rays\cite{auger23,rennie00,carniato20}, collisions with high-
energy particles such as electrons\cite{spohr70}, or electron 
capture\cite{loveland17}. They can also be specifically 
prepared and analyzed, which gives rise to site- and 
energy-specific spectroscopic methods to extract a variety of 
chemical information from molecules\cite{rye84,bolognesi12,
agarwal13,mcfarland14,ramasesha16,nisoli17,marchenko18,norman18,
kraus18,plekan20}, clusters\cite{tchaplyguine03}, and
materials\cite{harris68,hofmann12,agarwal13,zimmermann20} 
including surfaces\cite{weissman81,spanjaard85,powell90,
hofmann12,orvis19,li19} and nanostructures\cite{chao07,raman11}.
Auger electrons are also relevant for radiomedicinal 
approaches.\cite{kassis04,kassis05,buchegger06,ku19,howell20,
pirovano20,borbinha20,pirovano21} 

In these contexts, a specific interest lies in core-holes 
localized on atoms from the third or higher period of the 
periodic table because these exhibit more than one shell of core 
electrons.\cite{spanjaard85,tchaplyguine03,kassis04,kassis05,
buchegger06,chao07,raman11,hofmann12,agarwal13,ramasesha16,
loveland17,kraus18,orvis19,li19,ku19,pirovano20,borbinha20,
howell20,pirovano21} The different core electrons can be 
selectively ionized since they are well-separated in energy, 
which gives rises to specific Auger spectra. Further complexity 
is added because also the electrons involved in Auger electron 
can stem from different shells. In Auger decay, the respective 
spectra are dubbed according to the shells of the participating 
electrons. The first letter describes the initial hole, while 
the other two letters correspond to the final holes. A $KLL$ 
spectrum, for example, includes channels where a vacancy in the 
$K$ shell (1s) gets filled by an electron from the $L$ shell (2s 
and 2p) while an electron from the same shell gets emitted. 

In Coster-Kronig decay, one of the refilling electrons stems 
from the same shell as the initial core-hole\cite{coster35}. 
This manifests in the emission of very slow Auger electrons. An 
example for Coster-Kronig decay is an $LLM$ decay process where 
a 2p electron refills a 2s hole, and a electron from the $M$-
shell is emitted. Because of the high orbital overlap, this is 
much more efficient than intrashell Auger decay and one observes 
extraordinarily short lifetimes for s-type holes than for those 
of higher angular momenta within the same shell.\cite{bambynek72} These 
processes are to be distinguished from decay holes in an inner 
subshell of the outermost electron shell, which can only decay 
by ionization of the environment\cite{cederbaum97,zobeley01,
jahnke20,parravicini23}.

Auger decay involving other shells than the valence shell can 
leave the system with enough energy to undergo further Auger 
decay instead of vibrational or radiative relaxation. This 
behavior is typically referred to as a decay cascade and 
amplifies the number of Auger electrons produced from a single 
core-ionization, which is desirable for applications in 
radiotherapy\cite{howell20}. Decay cascades also leave 
signatures in Auger decay spectra when the spectra of the 
original core-ionized state and the primary decay products 
overlap.\cite{harris68}

Computational simulation of Auger decay is often a necessary 
supplement to experiments because it allows definitive assignments 
of signals to channels and electronic configurations.\cite{
fransson16,norman18,kraus18} An important feature of core-vacant 
states is that Auger decay makes them unbound. Such electronic 
resonances, i. e. all states which undergo autoionization, 
require special quantum-mechanical methods to describe the 
coupling to the continuum.\cite{moiseyev11,matz22,jagau22} 

One can distinguish between methods which only aim to compute 
the energy of Auger decay channels, with which a spectrum can 
only be constructed when assuming every decay channel to have an 
identical intensity, and methods where the decay channels' 
intensities are explicitly calculated which allows prediction of 
peak shapes in spectra. If only the energy of the core-ionized 
state is necessary, the core-valence separation (CVS) is often 
applicable and core-valence (Auger) transitions that make the 
system unbound are removed from the excitation manifold.\cite{
cederbaum80,wenzel14a,wenzel14b,coriani15,vidal19,liu19,bari19,
frati19,zheng19,nanda20,bokarev20,fransson21} This method 
reaches its limits where the energetic separation between the 
orbital group defined as core within these methods, where the initial hole is located and the one defined as valence, which is involved in Auger decay, is low, as 
for example in the presence of Coster-Kronig transitions. Also, 
assuming decay channels of equal intensity is a bad 
approximation in systems involving more than two shells as the 
intensities of these bands typically differ strongly.

To explicitly account for the decay process and compute 
probabilities for each of the decay channels (expressed as 
partial widths), different methods have been proposed, such as 
R-matrix theory\cite{gorczyca00,garcia09,tennyson10} or Fano's 
theory\cite{fano61,feshbach62,lowdin62,averbukh05,inhester12,
inhester14,kolorenc20,skomorowski21a,skomorowski21b,tenorio22}, 
which require explicit descriptions of the final states of the 
system and the emitted electron. We follow a different approach 
here, where the outgoing electron is rather described implicitly 
through complex scaling the coordinates in the Hamiltonian\cite{
aguilar71,balslev71,aguilar71,balslev71,moiseyev11} or the basis 
functions\cite{mccurdy78,moiseyev79} without a need to 
partition the Hilbert space. This scaling has the effect that 
the continuum wave functions become $L^2$ integrable and can 
therefore be treated by standard quantum chemical methods\cite{
bravaya13,jagau14,zuev14,white15a,white15b,white17,jagau17,matz22,
matz23a,matz23b,jayadev23}. The other effect is that the 
eigenenergies become complex, where the imaginary part relates 
to the decay width of the respective state which is inversely 
proportional to the life time. The partial widths are extracted 
via the decomposition of the complex coupled-cluster with 
singles and doubles (CCSD) energy\cite{matz22} or by applying 
Auger channel projectors on the equation-of-motion ionization 
potential (EOMIP)-CCSD wave function.\cite{matz23a}

Prior applications of the CCSD and EOMIP-CCSD methods, combined 
with complex scaling of the Hamiltonian or the basis functions, 
only considered atoms and molecules comprising elements from the 
first and second row like neon\cite{matz22,matz23a}, water\cite{matz22,matz23a}, molecular nitrogen\cite{matz22}, benzene\cite{matz22,jayadev23} or other hydrocarbons\cite{matz23a,matz23b}. Here, we apply this method to heavier 
elements. Our main focus laid on hydrogen sulfide because of its 
similarity to water and the exhaustive experimental and 
theoretical literature available.\cite{faegri77,puttner16,
asplund77,hikosaka04,poygin06,bueno03,svensson91,svensson94} We 
also investigated the argon atom as a simpler system which 
allows for a more thorough benchmarking. Both systems have five 
core orbitals (1s, 2s, $3\times$2p) which are atom-like and do 
not participate in bond formation. Thus, five different core 
holes exist for each of the systems which differ in decay width 
and spectral shape.

In addition to the computation of Auger decay widths, partial 
widths, and spectra, we compare the results to the water 
molecule regarding the spectral shape, the distribution of the 
decay among the decay channels and the relative contribution of 
triplet channels to the decay width. We also propose the use of 
the faster but less exact methods MP2 and EOMIP-MP2 which are 
based on a perturbational treatment of the correlation in the 
reference wave function.

This manuscript is structured as follows: In section II, the 
theory of complex scaling and complex basis functions, 
\mbox{(EOM-)MP2} and \mbox{(EOM-)CCSD}, and decomposition and 
Auger channel projectors will be discussed. We also explain how 
we obtain the positions and widths of the peaks in the spectra. 
In section III, we will introduce the systems of interest H$_2$S 
and Ar, before we discuss the computational details in section 
IV. In section V, we present the ionization energies, total Auger decay widths, and lastly the partial decay widths and some spectra. The manuscript will be 
concluded with an outlook in section VI.

\section{Theoretical description of Auger decay}
Core-vacant states are electronic resonances and thus not bound but metastable. They can undergo Auger decay and are therefore coupled to the continuum. Continuum wave-functions are not $L^2$ integrable, as they approach a plane wave at infinity, which poses a major challenge to standard computational methods. 

The approaches used throughout the article to describe the resonance character of core-ionized states are based on complex scaling or complex basis functions. Extensive discussion of the description of core-vacant states using these methods can be found elsewhere.\cite{matz22,matz23a,jagau22} Hereafter we delineate the aspects of the theory relevant to our work.

\subsection*{Complex scaling and complex basis functions}
Complex scaling\cite{aguilar71} (CS) of the Hamiltonian $\hat{H}$ describes the following similarity transformation
\begin{equation}
    \hat{H}_{\text{CS}}=\hat{S} \hat{H} \hat{S}^{-1}, \ \hat{S}=\text{e}^{\text{i} \theta r \frac{\text{d}}{\text{d}r}}, \ \theta \in \left(0, \frac{\pi}{4}\right).
\end{equation}

The effect of this on the energy of the continuum wave functions is a rotation with an angle $2\theta$ into the lower half of the complex plane. The resonance states become $L^2$ integrable, given $\theta$ is larger than a critical angle $\theta_{\text{c}}$ which is in the order of $0.01$° for Auger decay in neon.\cite{matz22} The eigenenergies of the resonances $E_{\text{res}}$ become complex due to the transformation
\begin{equation}
 E_{\text{res}}=E_{\text{R}} -\text{i} \frac{\Gamma}{2}.
\end{equation}
Here $E_{\text{R}}$ is the resonance position and $\Gamma$ the total decay width of that state, which is related to the lifetime $\tau$ by $\Gamma = \frac{\bar{h}}{\tau}$, where $\bar{h}$ is the reduced Planck's constant. When the Schrödinger equation is treated exactly, $E_{\text{res}}$ is independent on $\theta$ as long as $\theta > \theta_{\text{c}}$. Due to the finiteness of practical basis sets $E_{\text{res}}$
does depend on the scaling angle $\theta$. For this reason, one performs calculations for all possible angles, here in steps of 10~mrad, and minimizes $|\frac{\text{d} \Delta E}{\text{d}\theta}|$ because this derivative should be zero with a complete basis set. Here, $\Delta E$ is the energy difference of the ground and core-vacant state.

CS has the major flaw that it cannot be used to model molecular systems as it is not compatible with the Born-Oppenheimer approximation.\cite{matz22,jagau22} Therefore we can only use it for argon. To describe the Auger decay of H$_2$S we use the method of complex basis functions (CBFs)\cite{mccurdy78} which is based on the identity
\begin{align}
 \frac{\langle \psi (r) | \hat{H}(r \text{e}^{\text{i}\theta})| \psi (r)\rangle}{\langle \psi (r)| \psi (r)\rangle} = 
 \frac{\langle \psi (r \text{e}^{-\text{i}\theta}) | \hat{H}(r)| \psi (r \text{e}^{-\text{i}\theta})\rangle}{\langle \psi (r \text{e}^{-\text{i}\theta})| \psi (r \text{e}^{-\text{i}\theta})\rangle}.
\end{align} 

Here, we do not scale the Hamiltonian (left-hand side) but the coordinates of the basis functions (right-hand side). This has the same effect on the $L^2$ integrability of the resonance wave functions and the eigenenergies. To preserve the dilation analyticity, we only complex scale some additional diffuse basis functions to describe the emitted electron. To note that we add m complex-scaled shells of azimuthal quantum number l to our basis we add an appendage +m(l) to the name of our basis set. For example, when we add four shells of s, p, and d type, we write +4(spd).
In CBF calculations we optimize $\theta$ in steps of 1°.

\subsection*{(Equation of motion-)coupled cluster and second-order M\o ller-Plesset perturbation theories}
Coupled-cluster theory in the singlet and doubles approximation (CCSD) was used to study the electronic structure of the systems of interest.\cite{shavitt09} Two approaches were then used to describe the core-ionized states. In the first one, two subsequent complex energy CCSD calculations were separately done for the neutral and core-vacant state of the system.
We then take the difference of the real energies as the ionization energy and the difference of the imaginary part as the negative half of the decay width.\cite{matz22} The imaginary part of the ground state is a computational artifact, as for a complete basis set it should be zero. With this approach, we aim to subtract this nonphysical decay width from the total width.

The second approach used was equation-of-motion (EOM)\cite{stanton93,bartlett12,sneskov12} ionization potential (IP) CCSD. Here, one acts with an operator 
\begin{align}
 \hat{R}^{\text{IP}}= \sum_{i}^{\text{occ}} r_{i} i + \frac{1}{2} \sum_{ij}^{\text{occ}}\sum_{a}^{\text{virt}} r_{ij}^a a^{\dagger} ji
\end{align}
on the CCSD ket of the neutral molecule which introduces 1-hole ($i$) and 2-hole-1-particle ($a^{\dagger} ji$) excitations with amplitudes $r_{i}$ and $r_{ij}^a$, respectively. The core-hole state is then described as
\begin{align}
 |\Psi_{\text{EOM-IP-CCSD}}\rangle = \hat{R}^{\text{IP}}|\Psi_{\text{CCSD}}\rangle =\hat{R}^{\text{IP}} \text{e}^{\hat{T}}|\Phi_{HF}\rangle.
\end{align}
More details on the CCSD and EOM-CCSD approaches for the treatment of electronic resonances can be found elsewhere.\cite{shavitt09,jagau17,jagau22} 
Furthermore, in this work, we also explore the performance of second-order M\o ller-Plesset perturbation theory (MP2) in describing Auger decay for different core-holes. 
Analogous procedures to the ones presented for CCSD were carried out for the description of the core-holes within MP2.\cite{stanton95} 

\subsection*{Partial decay widths}
As in Auger decay, usually, a variety of final states are possible, we are interested in how the total decay width ($\Gamma$) splits up in these individual decay channels. Since the partial width divided by the total width equals the relative probability that the resonance decays via the respective channel, the partial decay widths can be used as a proxy for the intensity of Auger decay channels.
Two different methods have been proposed in the scope of the CS/CBF treatment of Auger decay.

The first approach is based on an energy decomposition analysis which only gave good results for CCSD and not for EOM-CCSD in previous projects.\cite{matz22} We therefore only use it for the CCSD and 
MP2 calculations. For a fixed initial core-hole $a$, the decay channel is defined by the two holes $i$ and $j$ in the occupied part of the final state. Note again that the decay width is twice the negative imaginary part of the complex energy. Thus, the partial decay width for CCSD is given by
\begin{align}
 \gamma_{ij}^{\text{CCSD}}= -2 \ \text{Im} \left(\sum_{b}^{\text{virt}} \left(2t_i^a t_j ^b +t_{ij}^{ab} \right) \langle ab | \, | ij \rangle \right).
\label{pwCCSD}
\end{align}
Here we sum over all virtual orbitals $b$ that describe the emitted electron. As the two-electron integrals $\langle ab | \, | ij \rangle$ are automatically computed in any CCSD and MP2 calculation the computation of the partial widths does not bring much additional computational cost.
From equation (\ref{pwCCSD}) it is possible to derive an expression for $\gamma_{ij}$ when MP2 is used instead of CCSD. The partial decay width of a particular channel is then given by
\begin{align}
 \gamma_{ij}^{\text{MP2}}= -2 \ \text{Im} \left(\sum_{b}^{\text{virt}} \frac{\langle ab||ij\rangle^2}{\epsilon_a +\epsilon_b -\epsilon_i -\epsilon_j}\right).
\end{align}

Within this approximation the $\Gamma_{ij}$ of a specific channel can be calculated without the need of perform a full energy calculation since only certain integrals are needed. This has the small caveat that the optimal $\theta$ still need to be optimized.

For the EOM-MP2 and EOM-CCSD calculations, the Auger channel projector (ACP) approach was used.\cite{matz23a}
Here one excludes the determinants corresponding to a specific Auger decay channel from the excitation manifold. By analyzing the effect of this on the complex energy in comparison to the full excitation manifold, one obtains the partial decay width of the respective channel. As this has to be repeated for every decay channel the ACP approach is usually computationally more expensive than the decomposition analysis. We therefore only use it for the EOM methods, where the decomposition method gave bad results in previous projects.\cite{matz22} 

\subsection*{Positions and broadening of the peaks in the spectra}
To simulate an Auger spectrum we need not only the partial widths (the height of the peaks) but also the broadening of the peaks and the kinetic energy of the respective emitted electron (the position of the peak). We calculate the latter by comparing the energy of the core-vacant state, which we get from a real EOM-IP-CCSD calculation, with the energy of the respective doubly ionized state. The latter is calculated by a real EOM double ionization potential (DIP-)CCSD calculation,\cite{sattelmeyer03,shen13,bokhan18} which is equivalent to EOM-IP-CCSD with the difference that the operator $\hat{R}$ now introduces 2-hole and 3-hole-1-particle excitations 
\begin{equation}
 \hat{R}^{\text{DIP}}= \frac{1}{2}\sum_{ij}^{\text{occ}} r_{ij} ji + \frac{1}{6} \sum_{ijk}^{\text{occ}}\sum_{a}^{\text{virt}} r_{ijk}^a a^{\dagger} kji.
\end{equation}
 The DIP states can be composed of multiple decay channels. We then assign to these states a sum of decay widths, where the weighting factors are given by the relative squared amplitudes of the channel to that state. This procedure has successfully been used in previous projects.\cite{jayadev23} All relevant DIP states can be found in the supplementary material. 

In experiments, the kinetic energy of emitted electrons cannot be measured perfectly. Different mechanisms lead to a broadening of the spectral lines like lifetime broadening, Doppler broadening, or pressure broadening. Depending on which mechanisms dominate, the lines show Gaussian (Doppler) or Lorentzian (lifetime and pressure) profiles. As the broadening, for some part, depends on the respective experiment, there is no \textit{a priori} optimal way to model the peak widths. We therefore adjust ourselves by the experiments that we want to compare our computations to and choose either Gaussian or Lorentzian and the full width at half maximum (FWHM) to match the experimental line shapes as well as possible.

\section{Electronic structure of hydrogen sulfide and argon}

Neutral hydrogen sulfide has 18 electrons, just like the noble gas argon. H$_2$S belongs to the C$_{2\text{v}}$ molecular point group, which has the four irreducible representations a$_1$, a$_2$, b$_1$, and b$_2$. Its electronic configuration is (1a$_1$)$^2$(2a$_1$)$^2$(1b$_1$)$^2$(3a$_1$)$^2$(1b$_2$)$^2$(4a$_1$)$^2$(2b$_1$)$^2$(5a$_1$)$^2$(2b$_2$)$^2$. Here, we use the Q-Chem symmetry notation instead of the traditional Mulliken's notation. The first five orbitals are effectively atom-like as they represent the sulfide's 1s, 2s, and 2p orbitals. Plots of all occupied orbitals can be found in the supplementary material. 

The argon atom also with 18 electrons belongs to the SO(3) molecular point group. Its electronic configuration is (1s)$^2$(2s)$^2$(2p)$^6$(3s)$^2$(3p)$^6$. All argon calculations were carried out in D$_{2h}$, the largest Abelian subgroup of SO(3). 

\section{Computation details}
The geometry of the H$_2$S molecule was optimized using the resolution-of-the-identity MP2 method\cite{weigend98}, the aug-cc-pCVTZ basis and the riMP2-aug-cc-pCVTZ auxiliary basis sets. We found a bond length of $1.334 \ \AA$ and an angle of $92.205$° in good agreement with the reported results.

All complex energy calculations of H$_2$S were carried out using the aug-cc-pCVTZ(5sp) basis for sulfur and aug-cc-pVTZ(5sp) basis for hydrogen. These basis sets were built by substituing the s and p exponents of the aug-cc-pCVTZ basis with those from the aug-cc-pCV5Z basis. This procedure has shown good results in previous projects.\cite{matz22,matz23a} The exponents of the added complex-scaled shells were calculated by scaling the optimized values to Ne\cite{matz23a} with respect to the geometric mean $\Bar{\zeta}$ of the exponents in the aug-cc-pCVTZ(5sp) basis set.\cite{matz22} We found $\Bar{\zeta} =5.50, 3.02$, and $0.42$ for Ne, S and H, respectively. Therefore we scale the exponents of Ne by a factor $0.550$ and $0.076$ for S and H, respectively. 

In calculations where we went beyond four complex-scaled shells per angular momentum, we produced additional exponents by adding even-tempered exponents to the basis set. Furthermore, when the initial core hole is in one of the three 2p-like orbitals, one should theoretically need f-type CBFs.\cite{chelkowska91} We obtained them for sulfur by taking the exponents from the d-type functions and scaling them with a factor 0.8 as this is also roughly the factor between d- and f-type exponents in the aug-cc-pVTZ(5sp) basis set. The final basis set can be found in the supplementary material.

In the case of argon, the full aug-cc-pCV5Z basis set was used in order to have enough an appropriate basis set for the complex scaling method. The exponents for the CBF calculations were calculated in the same way as for hydrogen and sulfur ($\Bar{\zeta}_{\text{Ar}}= 2.80$) and can also be found in the supplementary material.

The energy of the initial core-ionized and the final double ionized states were calculated using also the aug-cc-pVTZ(5sp) basis set. However, we were not able to converge any state that involved the 2a$_1^{-1}$ (2s$^{-1}$) hole of H$_2$S with a real EOM-IP or EOM-DIP calculation. For that reason, in the case of the IP energy of the 2s$^{-1}$ state we took the real part of the complex EOM-IP-CCSD energy. The missing energies of the doubly ionized states were calculated using an extrapolation procedure. We first computed the energy of all doubly ionized states with a reduced operator $\hat{R}^{\text{DIP, red}}=\frac{1}{2}\sum_{ij}^{\text{occ}} r_{ij} ji$ that only involves 2-hole excitations, which converges for every state. We then compare, for the states where the calculation with the non-reduced $\hat{R}^{\text{DIP}}$ converged, the obtained energies with full or reduced excitation manifold. Subsequently, we calculated the correlation energy 
as a function of the energy obtained with the $\hat{R}^{\text{DIP, red}}$ calculation. We linearly extrapolate this to get the approximate correlation energy for the missing states. The extrapolation procedure for all the missing energies can be found in the supplementary material. In the case of argon, it was only possible to converge some of the doubly ionized states that include a 2s$^{-1}$ hole. For the missing energies, we applied the same procedure as for H$_2$S.

All calculations were performed on a modified version of the Q-Chem program package.\cite{epifanovsky21}

\section{Results}
\subsection{Core ionization energies}

Table \ref{ioization enerues} shows the ionization energies (IEs) of the different core holes studied in this work. The given values are the real parts of the energies from complex EOM-IP calculations. We used for H$_2$O +2(spd), for H$_2$S 1s$^{-1}$ +4(spd), for for H$_2$S 2s$^{-1}$ +8(spd), for the H$_2$S 2p$^{-1}$ states +4(spdf) and for Ar 2s$^{-1}$ +8(spd).

\begin{table}[]
\caption{Ionization energies of the studied core holes.}\centering
\label{ioization enerues}
\begin{tabular}{ccc}
\hline
\textbf{Core hole} & \multicolumn{1}{l}{\textbf{Ionization energy/eV}} & \textbf{Experiment} \\ \hline
H$_2$O 1s$^{-1} $          &541.42
&     $538.728 \pm 0.017$\cite{wang2021calibration}                            \\
H$_2$S 1s$^{-1}$          &2475.75 
& $2478.25 \pm 0.40$\cite{puttner16}                                   \\
H$_2$S 2s$^{-1}$          & 234.99
&    $235 \pm 0.1$\cite{hikosaka04}                                \\
H$_2$S 1b$_1^{-1}$         & 171.95
&         Not measurable                         \\
H$_2$S 1b$_2^{-1}$         & 171.83
&           Not measurable                         \\
H$_2$S 3a$_1^{-1}$         & 171.78
&           Not measurable                        \\
Ar 1s$^{-1}$             & 3198.85
&         $3206.3 \pm 0.3$\cite{breinig80}                          \\
Ar 2s$^{-1}$             & 326.58
&      $326.25 \pm 0.05$\cite{glans93}                        \\\hline
\end{tabular}
\end{table}
The comparison of the theoretical IEs computed for the studied systems with the experimental data, when available, reveals that the biggest discrepancies occur for the 1s core holes. In the case of water, the energies are higher by $\sim$1 eV, while for H$_2$S and Ar, the differences are around 2.5 eV and 6.41 eV respectively. Probably, one of the causes of this increment is the lack of a proper treatment of the relativistic effects that start to be relevant in third-row elements.

\subsection{Total Auger decay widths}
\subsubsection*{K-edge Auger decay}
The 1s$^{-1}$ Auger decay in H$_2$S, was studied using all four methods previously discussed, MP2, CCSD, EOM-MP2, and EOM-CCSD. See table \ref{H2s 1s total widths} for the optimal $\theta$, the total widths, and the sum of the partial widths for the different methods and number of added complex-scaled shells. We find good agreement between the methods in the total width and the sum of partial widths with two complex-scaled shells per angular momentum except for EOM-CCSD. For the latter case, we went to four shells to get a better agreement with the other methods in the sum of partial widths. For the CCSD method, we found that going to four shells is not needed. Our calculated total width varies between 419 and 490 meV between the methods, whereas the sum of partial widths varies between 400 and 448 meV. There are multiple reasons for the discrepancy between total width and sum of partial widths.\cite{matz22,matz23a} Probably the most important is that the decomposition and ACP methods do not account for excited determinants that relate to shake-up and shake-off processes, which do contribute to the total width. The calculated values lay inside the error bars of the experimental value.\cite{keski74} The semi-empirical theoretical value from Krause and Oliver\cite{krause79} lays about 40$\%$ higher but is calculated with a much weaker theory. The theoretical value from Faegri and Keski-Rahkonen\cite{faegri77} is interpolated from a calculation from Chen and Crasemann\cite{chen73} but only includes the $KLL$ Auger decay. The $KLL$ part of the spectrum makes up between $87\%$ and $89\%$ of the total width in our calculations, dependent on the method. Our values for the $KLL$ part of the sum of partial width thus lie between 347 meV and 400 meV.

\begin{table}[t]
\caption{Optimal angle $\theta_{\text{opt}}$, total width, and the sum of partial widths (p.w.) of 1s$^{-1}$ Auger decay of hydrogen sulfide computed with different methods and number of complex-scaled shells.}\centering
\begin{tabular}{ccccc}
\hline
\textbf{Method} & \textbf{CBFs} &\textbf{$\theta_{\text{opt}}$} & \textbf{Total width} & \textbf{Sum of p.w.} \\ 
&&[°]&[meV]&[meV] \\
\hline

CCSD & +2(spd) & 30 & 443.6 & 446.6 \\

CCSD & +4(spd) & 14 & 436.1 & 451.8 \\

EOM-CCSD & +2(spd) & 40 & 419.2 & 276.0 \\

EOM-CCSD & +4(spd) & 16 & 430.6 & 399.8 \\

MP2 & +2(spd) & 40 & 484.1 & 448.1 \\

EOM-MP2 & +2(spd) & 19 & 490.7 & 430.7 \\

Theory\cite{krause79} & & & 590 & \\

Theory (only KLL)\cite{faegri77} & & & 430 & \\
Experiment\cite{keski74} & & & $500 \pm 100$ & \\ \hline
\end{tabular}
\label{H2s 1s total widths}
\end{table}

\subsubsection*{L$_1$-edge Auger decay}
\begin{table}[t]
\caption{Optimal angle $\theta_{\text{opt}}$, total width and sum of partial widths (p.w.) of the 2s$^{-1}$ Auger decay of hydrogen sulfide calculated with EOM-CCSD and +4(spd), +6(spd) and +8(spd).
}\centering
\begin{tabular}{ccccc}
\hline
\textbf{Method} & \textbf{CBFs} & $\theta_{\text{opt}}$° & \textbf{Total width} & \textbf{Sum of p.w.} \\
&&[°]&[meV]&[meV] \\
\hline
EOM-CCSD & +4(spd) & 17 & 1119.1 & 1020.0 \\
EOM-CCSD & +6(spd) & 27 & 1603.2 & 1407.4 \\
EOM-CCSD & +8(spd) & 14 & 1672.2 & 1440.5 \\
Experiment\cite{hikosaka04} & & & 1800 & \\
Theory\cite{krause79} & & & 1490 & \\ \hline
\end{tabular}
\label{H2S 2s total widths}
\end{table}

The description of the Auger decay process that occurs as a result of a vacancy in the 2a$_1$ orbital of hydrogen sulfide is more computationally challenging than that of the K-edge Auger decay. The reason for this is probably that the 2s$^{-1}$ state is much more correlated than the 1s$^{-1}$ state, due to the energetic proximity of the 2s orbital to the valence orbitals. Hence, CCSD does not converge in this case. Furthermore, MP2 gave very bad results as the total width increased from 0.526 to 19.5 eV, when going from two to five shells per angular momentum. Therefore we only consider calculations with EOM-CCSD.  We performed calculations with four, six, and eight complex-scaled shells per angular momentum. The first addition of CBFs increased the total width considerably and led to a better agreement with an experimental and another theoretical value, as can be seen in Table \ref{H2S 2s total widths}. Using eight instead of six shells increased the decay width again, but had a much smaller impact. The need for more diffuse CBFs can be explained by the low energy of the emitted electrons. Additionally, we observed that EOM-MP2 gives very similar results with almost no computational benefit.

The total width is almost four times as large as for the 1s hole even though fewer decay channels exist. The reason for that is the existence of very intense Coster-Kronig\cite{coster35} transitions where an electron of the same shell fills the hole and a valence electron with relatively little kinetic energy gets emitted. Here, a 2p electron fills the 2s$^{-1}$ hole, which are the $LLM$ transitions. Those decays happen so fast because the 2s and 2p orbitals have very large radial overlap.\cite{weightman18} The $LLM$ part makes up 1396 of the total 1441 meV (sum of partial widths with +8(spd)). Interestingly, triplet channels contribute much more to the total width here than for the 1s$^{-1}$ hole. The ratio between singlet and triplet channels is $0.76:0.24$, whereas for 1s$^{-1}$ it is $0.95:0.05$.

\begin{table}[t]
\caption{Optimal angle $\theta_{\text{opt}}$, total width, and the sum of partial widths (p.w.) of the 2s$^{-1}$ Auger decay of argon.}\centering
\begin{tabular}{ccccc}
\hline
\textbf{Method} & \textbf{CBF/Basis} & \textbf{$\theta_{\text{opt}}$} & \textbf{Total width} & \textbf{Sum of p.w.} \\ 
&&[°]&[meV]&[meV] \\
\hline
CS-EOM-CCSD & 5Z & 17° & 2632.3 & 2450.2 \\
CS-EOM-CCSD & QZ & 16° & 2531.8 & 2347.3 \\
CS-EOM-MP2 & 5Z & 16° & 2620.8 & 2382.0 \\
CBF-EOM-CCSD & +4(spd) & 14° & 1053.8 & 872.2 \\
CBF-EOM-CCSD & +6(spd) & 32° & 2100.9 & 2294.6 \\
CBF-EOM-CCSD & +8(spd) & 34° & 2668.6 & 2334.2 \\
Experiment\cite{glans93} & & & $2250 \pm 150$ & \\
Experiment\cite{glans93} & & & $2250 \pm 50$ & \\
Experiment\cite{mehlhorn66} & & & $1840 \pm 200$ & \\
Theory\cite{glans93} & & & 1850.0 & \\

Theory (only LLM)\cite{kylli99} & & & 2330.0 & \\
Theory\cite{krause79} & & & 1630 & \\ \hline
\end{tabular}
\label{Argon 2s}
\end{table}

The 2s$^{-1}$ Auger decay of argon has been investigated much more\cite{glans93,mehlhorn66,lablanquie11,ogurtsov69,kylli99,lablanquie00,avaldi19,pedersen23}than in the case of H$_2$S which enables a better comparison for our simulation. An advantage of argon is that we can use complex scaling which is more straightforward as it does not require the optimization of the CBFs' exponents and enables the comparison and therefore verification of the CBF approach.
Table \ref{Argon 2s} shows $\theta_{\text{opt}}$, the total width and the sum of partial widths calculated with EOM-CCSD and four, six and eight complex-scaled shells for s, p and d angular momentum, and complex scaling with aug-cc-pCV(Q and 5)Z (EOM-CCSD) and aug-cc-pCV5Z (EOM-MP2). We find that aug-cc-pCVQZ gives a very similar result to the more complete 5Z basis. Also, EOM-MP2 agrees very well with EOM-CCSD but brings almost no time benefit. Furthermore, we find that adding 4(spd) is not enough to describe the process properly. As for H$_2$S, we see that eight complex scaled shells only slightly change the sum of partial widths compared to six. The agreement between the CS and CBFs calculations is also very good, verifying our calculations for hydrogen sulfide. The values of our best calculated sums of partial widths 2450 meV (CS) and 2334 meV (CBF) also agree very well with the most recent experiment\cite{glans93} (2250 meV) and theoretical value\cite{kylli99} (2330 meV). The latter only describes the $LLM$ part of the spectrum, which contributes 97\% of the total sum of widths in our calculations. The older result for the total width is about 20\% smaller.

\subsubsection*{L$_{2,3}$-edge Auger decay}
\begin{table}[t]
\caption{Total width and the sum of partial widths (p.w.) in meV for the three 2p holes 1b$_2^{-1}$, 3a$_1^{-1}$ and 1b$_1^{-1}$ and the mixed states $3e_{1/2}^{-1}$, $4e_{1/2}^{-1}$, and $5e_{1/2}^{-1}$ of hydrogen sulfide.}\centering
\begin{tabular}{cccccccc}
\hline
\textbf{Initial hole} & \textbf{1b$_2^{-1}$}$^b$ & \textbf{3a$_2^{-1}$}$^b$ & \textbf{1b$_1^{-1}$}$^b$ & 
\textbf{5e$_{1/2}^{-1}$} & \textbf{2p$^{-1}$$^c$} \\ \hline
\textbf{$\epsilon_{1}^{-1}$} & \textbf{$\epsilon_{2}^{-1}$} & \textbf{$\epsilon_{3}^{-1}$} & \textbf{2p$^{-1}$$^c$} \\ \hline
Total width$^a$ & 86.1 & 76.7 & 68.6 & 76.3 & 72.4 & 82.8 & 77.1 \\
Sum of p.w.$^a$ & 52.4 & 47.1 & 38.7 & 45.4 & 42.6 & 50.2 & 46.0 \\
Experiment\cite{poygin06} & & & & $63 \pm 1$ & $65 \pm 5$ & $75 \pm 1$ & $68 \pm 2$ \\
Experiment\cite{bueno03} & & & & $64 \pm 2$ & & $74 \pm 2$ & $69 \pm 2 $ \\
Theory\cite{bueno03} & 90.5 & 69.8 & 49.7 & 68 & 59 & 83 & 70 \\
Theory\cite{krause79} & & & & & & & 54 \\ \hline
\multicolumn{8}{l}{$^a$Computed with EOM-CCSD/aug-cc-pCVTZ(5sp)+4(spdf).}\\
\multicolumn{8}{l}{$^b$$\theta_{\text{opt}}$ is 11°, 10° and 10° for 1b$_2^{-1}$, 3a$_1^{-1}$ and 1b$_1^{-1}$, respectively.}\\
\multicolumn{8}{l}{$^c$The width of 2p$^{-1}$ is taken as the average over the three holes.}
\end{tabular}
\label{H2S 2p table}
\end{table}

Auger electrons can also be produced when the core hole is present in one of the three 2p-like orbitals of H$_2$S, namely 1b$_1^{-1}$, 3a$_1^{-1}$, and 1b$_2^{-1}$. Again like for the 2s$^{-1}$ hole, the used solver for the CCSD equations does not converge and MP2 does not give trustful results. EOM-MP2 again shows almost identical results to EOM-CCSD but does not have much computational benefit. We therefore limit ourselves to the latter method here. There are now three possible holes (1b$_1^{-1}$, 3a$_1^{-1}$ and 1b$_2^{-1}$) that are almost energetically degenerate (the ionization potentials vary by about 0.1 eV, see table \ref{ioization enerues}).

As mentioned in section III, f-type added shells should now be necessary to describe the Auger decay. We performed multiple calculations for the three holes, with two and four complex-scaled shells per angular momentum and with or without complex-scaled f shells on sulfur. We found that the results with four shells are much better than with two, as we already saw for the 1s$^{-1}$ hole. The effect of the added f-type shells is, however, small. The effect may have been larger if we had also added f CBFs to hydrogen and not only to the sulfur basis, although it helped to find a clearer and more pronounced optimal $\theta$ in the optimization trajectory. We therefore present the results for this more complete basis. A comparison between the calculations with different methods and basis sets for the 1b$_2^{-1}$ hole can be found in the supplementary material.

Due to spin-orbit coupling the three 2p orbitals mix to 2p$_{1/2}$ and 2p$_{3/2}$ states. The ligand-field effect (the effect of the hydrogen atoms on the core structure of sulfur) splits the latter again. We will call the resulting states $\epsilon_{2,3}$
while the 2p$_{1/2}$ states will then be called
$\epsilon_{1}$.The latter is energetically separated by about 1.2 eV from the other two levels which are split up by about 110 meV.\cite{poygin06,svensson91} As these energy differences are on the order of magnitude of the lifetime broadening of the respective states, the various hole states overlap leading to lifetime interferences. The ionization that prepares the molecule in the core-vacant state and the Auger decay should therefore be described as a one-step process\cite{poygin06}, which is not done here. The mixing of the 1b$_1$, 3a$_1$ and 1b$_2$ orbitals into these
$\epsilon$ states have been previously reported\cite{bueno03} and can be found in the supplementary material. 

Therefore, we were able to compute the decay widths for both types of state descriptions.
This is shown in table \ref{H2S 2p table}, in comparison with experimental\cite{poygin06,bueno03} and other theoretical results\cite{krause79,bueno03}. We first observe that the width of the 2p$^{-1}$ state is one order of magnitude smaller than the 1s$^{-1}$ hole and almost two orders smaller than for the 2s$^{-1}$ hole. Note however that the $L_2MM$ width of H$_2$S with 46 meV is much larger than the $KMM$ width (1.3 meV) and of similar size as the $L_1MM$ width (44 meV). The reason is that the 2s and 2p orbitals have much more spatial overlap with the valence electrons than the 1s orbital. Furthermore, the discrepancy between our calculated total width (difference between the imaginary part of the energy of neutral ground state and core-vacant state) and the sum of partial widths obtained by the ACP procedure is more pronounced on a relative scale than for our previous calculations. The reason behind this is not clear. Although our calculated total widths are too large and our sum of partial widths too small in comparison to the experiments, our results agree well with the experiment that the
$\epsilon_3^{-1}$ state, which mostly consists of the 1b$_2^{-1}$ state, has the largest width. We also reproduce previous theoretical results which report that the width of 3a$_1^{-1}$ is larger than that of 1b$_1^{-1}$.\cite{bueno03}

\subsection{Auger Spectra}
\subsubsection*{K-edge Auger spectra}
\begin{figure}
\centering
\includegraphics[width=0.8\linewidth]{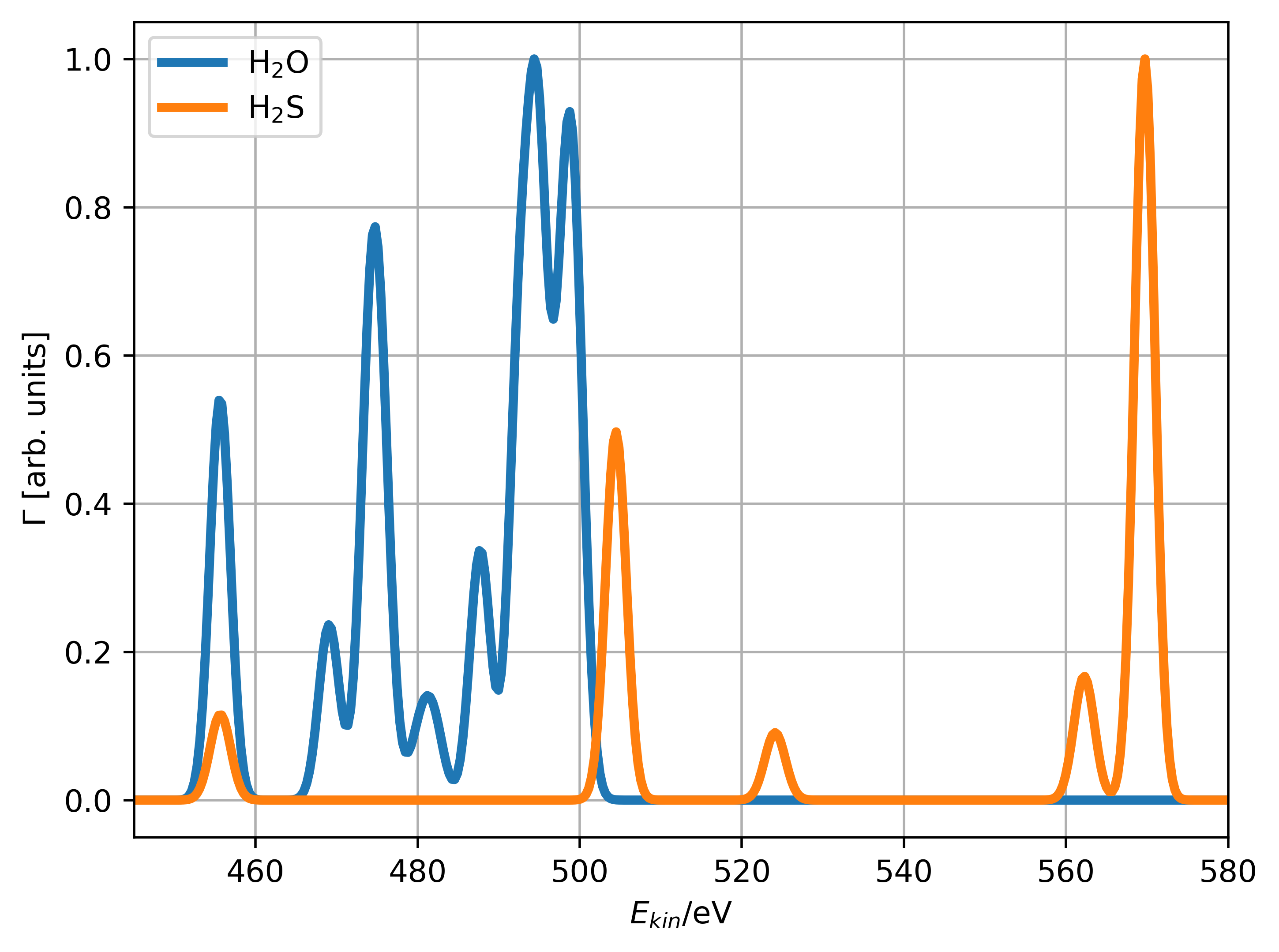}
\caption{Comparison of the $KLL$ Auger spectrum of water and hydrogen sulfide. Decay width
in arbitrary units as function of the kinetic energy of the emitted electron. The partial widths were calculated with CCSD and +2(spd). The peaks corresponding to the KLL Auger spectrum of H$_2$S were shifted 1515 meV. All peaks were normalized to ease the comparison. Gaussian broadening with FWHM = 3 eV.}
\label{KLL water}
\end{figure}

K-edge Auger spectra have been previously studied for different molecules using the discussed methodology.\cite{matz22,matz23a, skomorowski21a} In particular, the Auger decay of water, an example that has been extensively discussed, is an interesting point of comparison for the KLL Auger spectrum of H$_2$S due to the chemical and electronic similarities between these two molecules.
Figure \ref{KLL water} illustrates this comparison. Here we have shifted the KLL spectrum of H$_2$S by 1515 meV to match the lowest energy peak in the spectrum and ease the comparison. All the peaks in the KLL Auger spectrum of water can be found in an energy range of less than 50 eV while for H$_2$S this energy range is $\sim$130 eV, which can probably be explained by the difference in the energies between the orbitals in this region. Another significant difference between these two spectra are the number of peaks present in each of them. The Auger spectrum of water shows 7 peaks while in hydrogen sulfide only 5 peaks are visible. This difference can be caused by the fact that, in water, the decay occurs from valence orbitals, while in H$_2$S the orbitals involved retain a certain atomic character which is reflected in the spectrum and the partial widths.

From this figure it becomes clear that spite the chemical resemblances between H$_2$O and H$_2$S their Auger decay processes are very different. This conclusion can be further supported by looking at the partial widths for the 16 decay channels of the $KLL$ region for these two molecules. These results are presented in Table \ref{KLL pw H2O vs H2S}. By looking at the $KLL$ partial widths we observe that channels in H$_2$S are more intense that their counterpart in water with the exception of the three 3a$_1$—1b$_1$, 3a$_1$—1b$_2$, and 1b$_1$—1b$_2$ triplet channels. The relative intensities reveal that in both molecules the 3a$_1$—1b$_1$ singlet channel is the most intense of the $KLL$ spectrum. However, the relative intensities of the remaining channels exhibit significant variations, notably the pronounced intensity of the 2a$_1$—2a$_1$ channel in water, a feature that is less prominent in H$_2$S.

\begin{table}[t]
\caption{KLL partial decay widths in meV of water and hydrogen sulfide molecules.}\centering
\begin{tabular}{@{\extracolsep{8pt}}ccccc}
\hline
\multirow{3}{*}{\textbf{Decay channels}} & \multicolumn{2}{c}{H$_2$O} & \multicolumn{2}{c}{H$_2$S}                                                                      \\ \cline{2-3} \cline{4-5} 
& \multicolumn{1}{l}{\textbf{Partial}} & \multicolumn{1}{l}{\textbf{Relative}} & \multicolumn{1}{l}{\textbf{Partial}} & \multicolumn{1}{l}{\textbf{Relative}} \\ 

& \multicolumn{1}{l}{\textbf{width}} & \multicolumn{1}{l}{\textbf{intensity$^a$}} & \multicolumn{1}{l}{\textbf{width}} & \multicolumn{1}{l}{\textbf{intensity$^a$}} \\
\hline
2a$_1$ — 2a$_1$ (S) & 16.72 & 0.96 & 24.60 & 0.51 \\
2a$_1$ — 3a$_1$ (S) & 13.48 & 0.77 & 35.30 & 0.74 \\
2a$_1$ — 1b$_1$ (S) & 7.25 & 0.41  & 35.30 & 0.74 \\
2a$_1$ — 1b$_2$ (S) & 12.34 & 0.71 & 35.40 & 0.74 \\
2a$_1$ — 3a$_1$ (T) & 2.44 & 0.14 & 6.50 & 0.14 \\
2a$_1$ — 1b$_1$ (T) & 1.91 & 0.11 & 6.50 & 0.14 \\
2a$_1$ — 1b$_2$ (T) & 2.94 & 0.17 & 6.50 & 0.14 \\
3a$_1$ — 3a$_1$ (S) & 11.65 & 0.67 & 35.70 & 0.75 \\
1b$_1$ — 1b$_1$ (S) & 10.05 & 0.57 & 35.50 & 0.74 \\
1b$_2$ — 1b$_2$ (S) & 16.49 & 0.94 & 36.60 & 0.77 \\
3a$_1$ — 1b$_1$ (S) & 17.49 & 1.00 & 47.80  & 1.00 \\
3a$_1$ — 1b$_2$ (S) & 16.43 & 0.94 & 47.00 & 0.98 \\
1b$_1$ — 1b$_2$ (S) & 14.18 & 0.81 & 46.90 & 0.98 \\
3a$_1$ — 1b$_1$ (T) & 0.16 & 0.01 & 0.00 & 0.00 \\
3a$_1$ — 1b$_2$ (T) & 0.27 & 0.02 & 0.00 & 0.00\\
1b$_1$ — 1b$_2$ (T) & 0.00 & 0.00 & 0.00 & 0.00\\ \hline
\multicolumn{5}{l}{$^a$Calculated with respect to the most intense width in the region.}
\end{tabular}
\label{KLL pw H2O vs H2S}
\end{table}

The widths of all 64 decay channels of H$_2$S computed with the four methods discussed in this work can be found in the supplementary material. The analysis of these partial widths for the 1s$^{-1}$ Auger decay of H$_2$S shows a good agreement between the different methods, but only when using +4(spd) for EOM-CCSD. The agreement between CCSD and MP2 is surprisingly good. The widths of the significant channels only vary between about 1$\%$ and 5$\%$. The mean difference between these methods for the channels with a width larger than 1 meV is only 3\%. 

\begin{figure}
\centering
\includegraphics[width=0.8\linewidth]{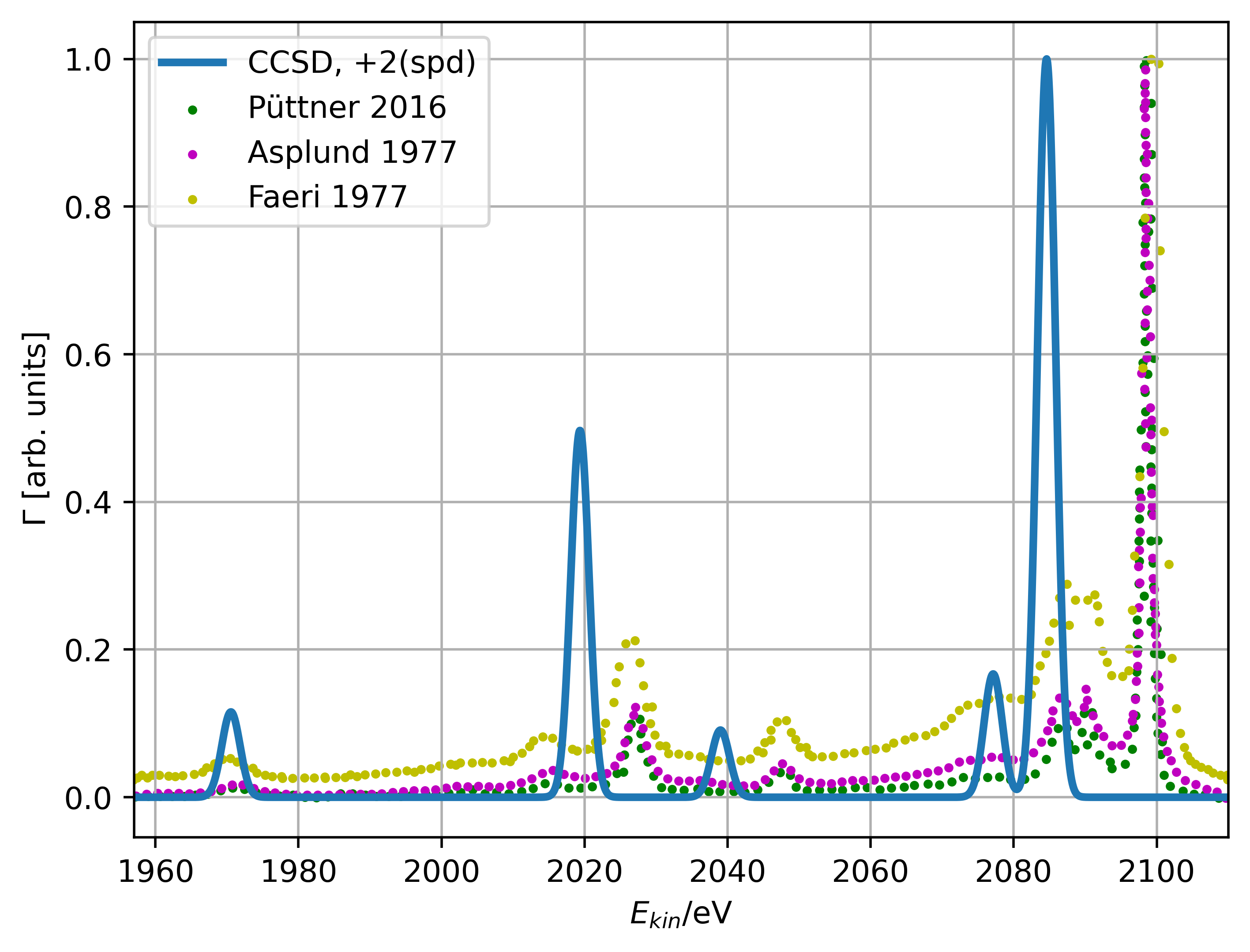}
\caption{$KLL$ Auger spectrum of hydrogen sulfide. Decay width
in arbitrary units as function of the kinetic energy of the emitted electron. The blue line shows our result. The partial widths were calculated with CCSD and +2(spd).
Gaussian broadening with FWHM = 3 eV.}
\label{KLL H2S}
\end{figure}

We will keep focusing our attention on the $KLL$ part of the spectrum and compare it with three experimental measurements.\cite{faegri77,puttner16,asplund77} This is shown in figure \ref{KLL H2S}. We observe five peaks corresponding (from left to right) to the 2s$_1^{-2}$ ($^1$S), 2s$^{-1}$ 2p$^{-1}$ ($^1$P), 2s$^{-1}$ 2p$^{-1}$ ($^3$P), 2p$^{-2}$ ($^1$S) and 2p$^{-2}$ ($^1$D) final states, respectively. The intensity in the spectra is normalized to 1 for the highest peak. We only show the CCSD result as a proxy, as all our four methods give similar spectra. Whereas the positions of the peaks agree very well in the experiments, there is some disagreement in our calculation. The highest peak (2p$^{-2}$ ($^1$D)), for example, lays 13.7 eV higher in energy for the experiment than in our computation. A reason could be that we did not include relativistic effects and spin-orbit splitting in our EOM-CCSD calculations. The major disagreement in the height of the peaks between the experiments and our calculations is that we overestimate the three peaks at lower kinetic energy compared to the other two. The overestimation is the largest for the CCSD and MP2 calculations. An overestimation of peaks at lower kinetic energies was already observed in a previous project with small molecules.\cite{matz23b}

A usual comparison of computed Auger spectra is with the equal width approximation. Here, one assumes that the width equally splits up into all possible channels. The corresponding plot can be found in the supplementary material. The agreement between our calculation and the experiments is very poor. The largest disagreement is that in the equal width approximation, an additional peak to the right of the large 2p$^{-2}$ arises stemming from a $^3$P channel that is forbidden and exactly zero in our calculations. In the experiments, this peak is finite but still very small due to spin-orbit coupling.\cite{puttner16} Lastly, note that except from the 2s$^{-1}$ 2p$^{-1}$ ($^3$P) state all triplet channels are negligible in the 1s$^{-1}$ calculation. The singlet channels contribute 425 meV of the total sum of widths of 447 meV in the CCSD calculation. This is very similar to the other methods.

While the sum of $KLL$ widths is 400 meV (for our CCSD result), the $KLM$ ($M$ = valence) and $KMM$ contribute 45.7 and 1.3 meV, respectively. The reason that these widths are smaller is that the $M$ electrons have a smaller spatial overlap with the $K$ hole than the $L$ electrons. The respective computed spectra can be found in the supplementary material. To the author's knowledge, there are no experimental results published of those spectra.

\subsubsection*{L$_1$-edge Auger spectra
}

\begin{figure}
\centering
\includegraphics[width=0.8\linewidth]{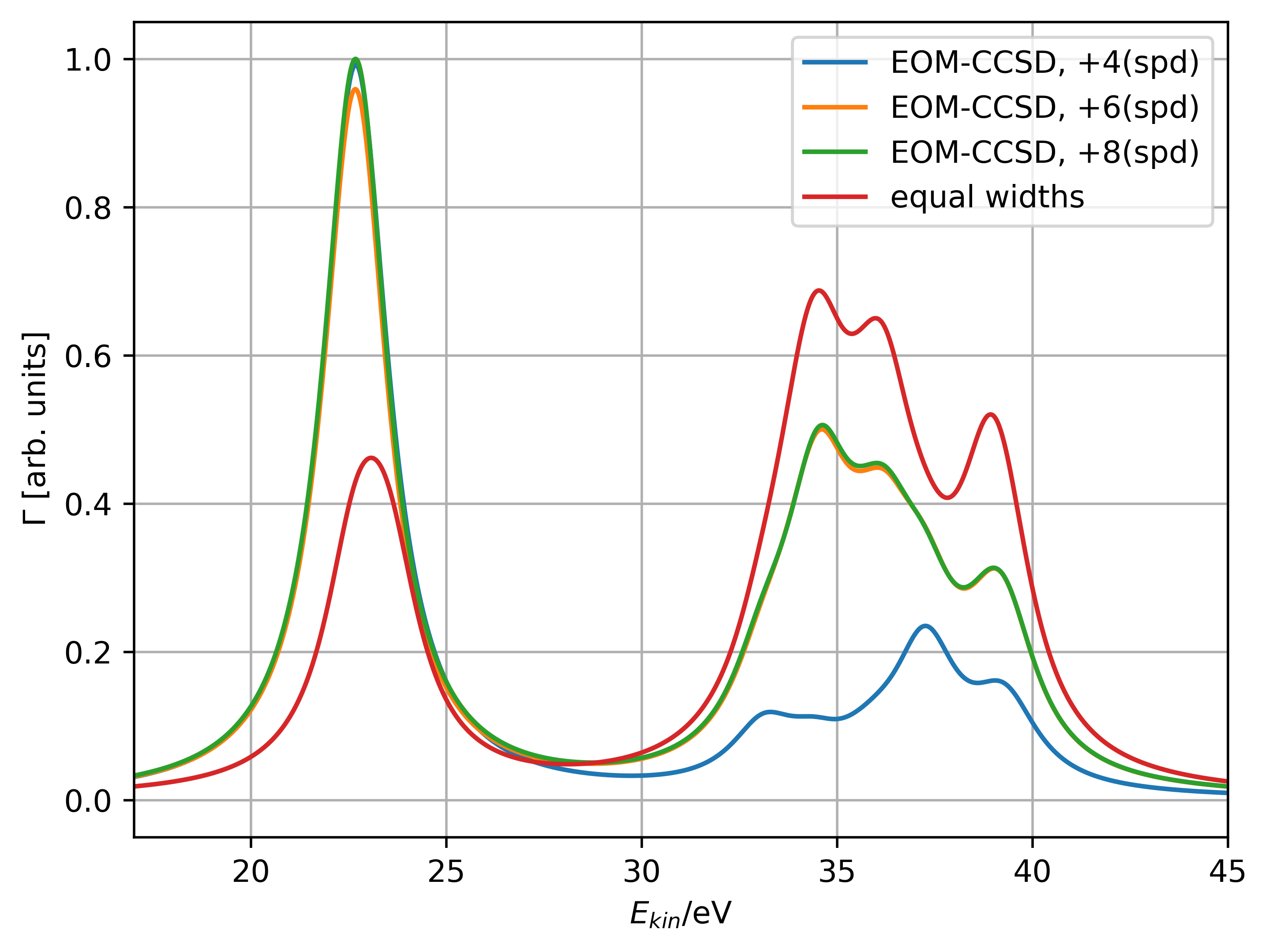}
\caption{$LLM$ Coster-Kronig spectrum of hydrogen sulfide. Decay width
in arbitrary units as a function of the kinetic energy of the emitted electron.
The partial widths were calculated with CBF-EOM-CCSD. The equal width result is normalized to the same sum of widths as the EOM-CCSD +8(spd) curve. 
Lorentzian broadening with FWHM = 2 eV.}
\label{LLM H2S}
\end{figure}

Figure \ref{LLM H2S} shows the $LLM$ Auger spectrum starting from the 2a$_1^{-1}$ core-ionized state of H$_2$S. This spectrum consists of 24 of the total 40 decay channels, and it contributes by far the most to the total width. We compare the EOM-CCSD with four, six, and eight complex-scaled shells with the equal widths result. We see two clearly separated structures in the spectrum. Towards lower kinetic energies at about 22.5 eV we see a singular peak which corresponds to the 2p$^{-1}$4$a_1^{-1}$ final states ($L_1L_2M_1$). In our calculations, the corresponding singlet channels are the most intense with 230 meV for the calculation with +8(spd). Note that our calculation produces negative decay widths ($\mathtt{\sim}$ -10 meV) for the corresponding triplet channels. The reason for this is not totally clear, however, it is certain that our theoretical model misses the spin-orbit coupling that leads to a splitting of the $^3$P level. The latter lay at a slightly higher energy than the corresponding singlet channel. This is the reason why the peak of equal widths is broader and seems to be shifted to the right.

Towards larger kinetic energies we see a more complex structure consisting of the 18 2p$^{-1}$$M_2^{-1}$ (with $M_2$= 2b$_1$,5a$_1$,2b$_1$) final states ($L_1L_2M_2$). The calculation with four complex shells here also gives some negative widths, which is not the case for six and eight added complex-scaled shells. Our calculated ratio between $L_1L_2M_1$ and $L_1L_2M_2$ is $0.47:0.53$ ($\frac{6}{24}:\frac{18}{24}$ for equal widths) for the +8(spd) result. Unfortunately, to our knowledge, an experimental measurement of the $LLM$ spectrum of hydrogen sulfide with sufficient resolution that allows a meaningful comparison with our simulation has not been reported. However, a very noisy $L_1L_2M_2$ spectrum was reported, where three peaks are weakly visible which agrees with our calculations.\cite{hikosaka04}

The fact that we find some negative decay widths and only poor agreement between the EOM-CCSD result with four complex shells on the one hand, and six and eight complex-scaled shells on the other side, for the $L_1L_2M_2$ part can at least partly be explained by the exponents of the complex shells being not optimized to describe the slow emitted electrons.

\begin{figure}
 \centering
 \includegraphics[width=0.8\linewidth]{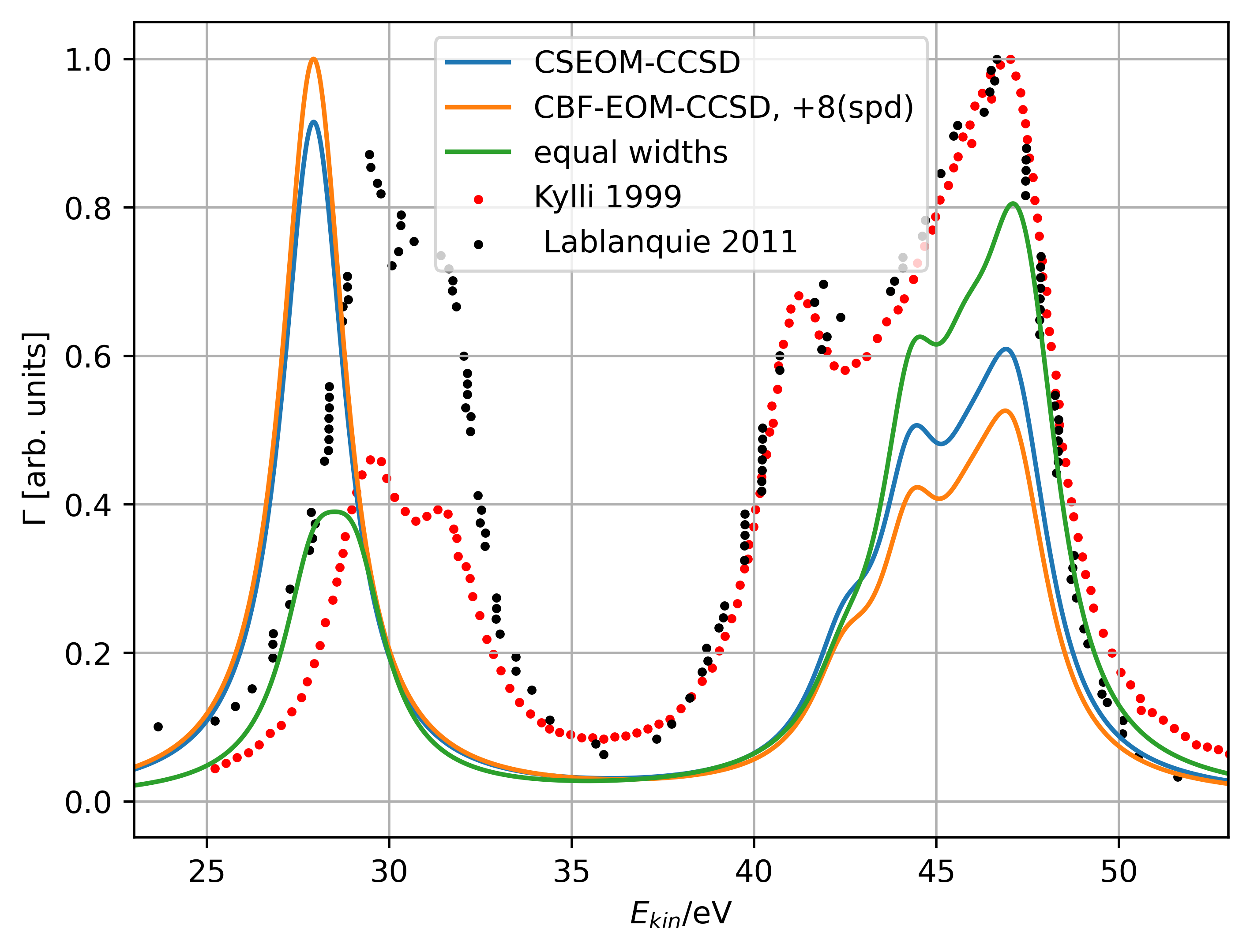}
 \caption{$LLM$ Coster-Kronig spectrum of argon. Decay width
 in arbitrary units as a function of the kinetic energy of the emitted electron. The partial widths were calculated with CS-EOM-CCSD (blue), CBF-EOM-CCSD,
 and equal widths (green). 
 The calculated spectra are shifted by 3.4 eV towards higher energies to align with the experiments on the rightmost peak. 
 The spectra are normalized to ease the comparison.
 Lorentzian broadening with FWHM = 2 eV.}
 \label{LLM ar}
\end{figure}

The 2s$^{-1}$ Auger decay of argon\cite{glans93,mehlhorn66,lablanquie11,ogurtsov69,kylli99,lablanquie00,avaldi19,pedersen23} has received considerably more attention than H$_2$S which enables better point of comparison for our methodology. As argon and hydrogen sulfide have a similar core structure the simulated spectra look very similar for the $LLM$ part. We can therefore draw conclusions on our H$_2$S results by comparing our argon calculations with the experiment. The other advantage of argon is that we can use complex scaling which is more straightforward as it does not require the optimization of the CBFs' exponents and enables the comparison and therefore verification of the CBF approach.
The partial widths can again be found in the supplementary material. We here focus on investigating the $LLM$ spectrum calculated with CS and CBFs which is shown in figure \ref{LLM ar}. For comparison, an equal width result, which is normalized to the same total width as the CBFs calculation, and two experimental spectra are given. The calculated spectra are shifted to the right such that the rightmost peak aligns with the experiments that are normalized to unity at their highest peak. We first observe the reasonable agreement between the CS and CBFs results. The main difference lies in the partition of the total width between the $L_1L_2M_1$ (around 25 eV) and $L_1L_2M_2$ (around 45 eV). The $L_1L_2M_1$:$L_1L_2M_2$ splitting is 0.44:0.56 for CS and 0.50:0.50 for CBF, which are similar to the ratio in H$_2$S. The experimental values for this division are 0.21:0.79\cite{mehlhorn66} and 0.23:0.77\cite{kylli99}. The CS result is in this sense slightly better than CBF. The discrepancy between experiment and calculation for this ratio is long known and have been previously discussed.\cite{kylli99,karim85} The authors also observed discrepancies in the $L_1L_2M_1$:$L_1L_2M_2$ ratio between experiment and theory in the order of a factor 2. They gave multiple possible reasons but most of their computational shortcomings (like electron correlation) are properly accounted for in our calculation. The other reason they give is spin-orbit coupling and the resulting mixing of decay channels. As spin-orbit interaction is completely missing in our calculation, we assume that this is the main reason for the mismatch with experiments. Note, however, that the two shown experiments\cite{kylli99,lablanquie11} in figure \ref{LLM ar} themselves also disagree in the $L_1L_2M_1$:$L_1L_2M_2$ ratio.

Furthermore, we see that the $L_1L_2M_1$ part shows a single peak in our calculation. The reason for this is that our calculations give negligible (CS) or even negative ($\mathtt{\sim}$ -5 meV, CBFs) decay widths for the corresponding triplet channels. This also has a very poor agreement with the experiments that show two peaks, a larger one from the singlet channels and a smaller one from the triplets. The missing of the triplet channels in simulation was also intensively discussed before. The main reason is channel mixing due to spin-orbit coupling which can change the singlet to triplet ratio by as much as a factor 84.\cite{karim85,karim84,bruneau83}

We again see also a disagreement in the positions of the peaks compared to the experiment. This is also tied to the missing spin-orbit coupling in the real EOM-IP-CCSD and EOM-DIP-CCSD calculations.

In addition, note that the effect of spin-orbit coupling and channel mixing is not limited to argon or even the Coster-Kronig transitions. We expect that the same effects impair our other results as well. For the 2s$^{-1}$ spectrum of hydrogen sulfide, the $L_1L_2M_1$ is most likely also overestimated compared to $L_1L_2M_2$. The missing triplet peak in the $L_1L_2M_1$ part should also be included when we add spin-orbit coupling. It was also reported that $KLL$ spectra are affected by channel mixing, too, though to a smaller degree.\cite{howat78} We, therefore, expect a better agreement of our $KLL$ spectrum in the position and heights of the peaks with experiments as well when our theory accounts for spin-orbit interaction.

Lastly, channel mixing and spin-orbit splitting almost exclusively lead to a redistribution of the partial widths but not to a change of the total width.\cite{karim85} This is the reason that our total width agreed very well with experiments for argon and hydrogen sulfide.

\subsubsection*{L$_{2,3}$-edge Auger spectra}
\begin{figure}
\centering
\includegraphics[width=0.8\linewidth]{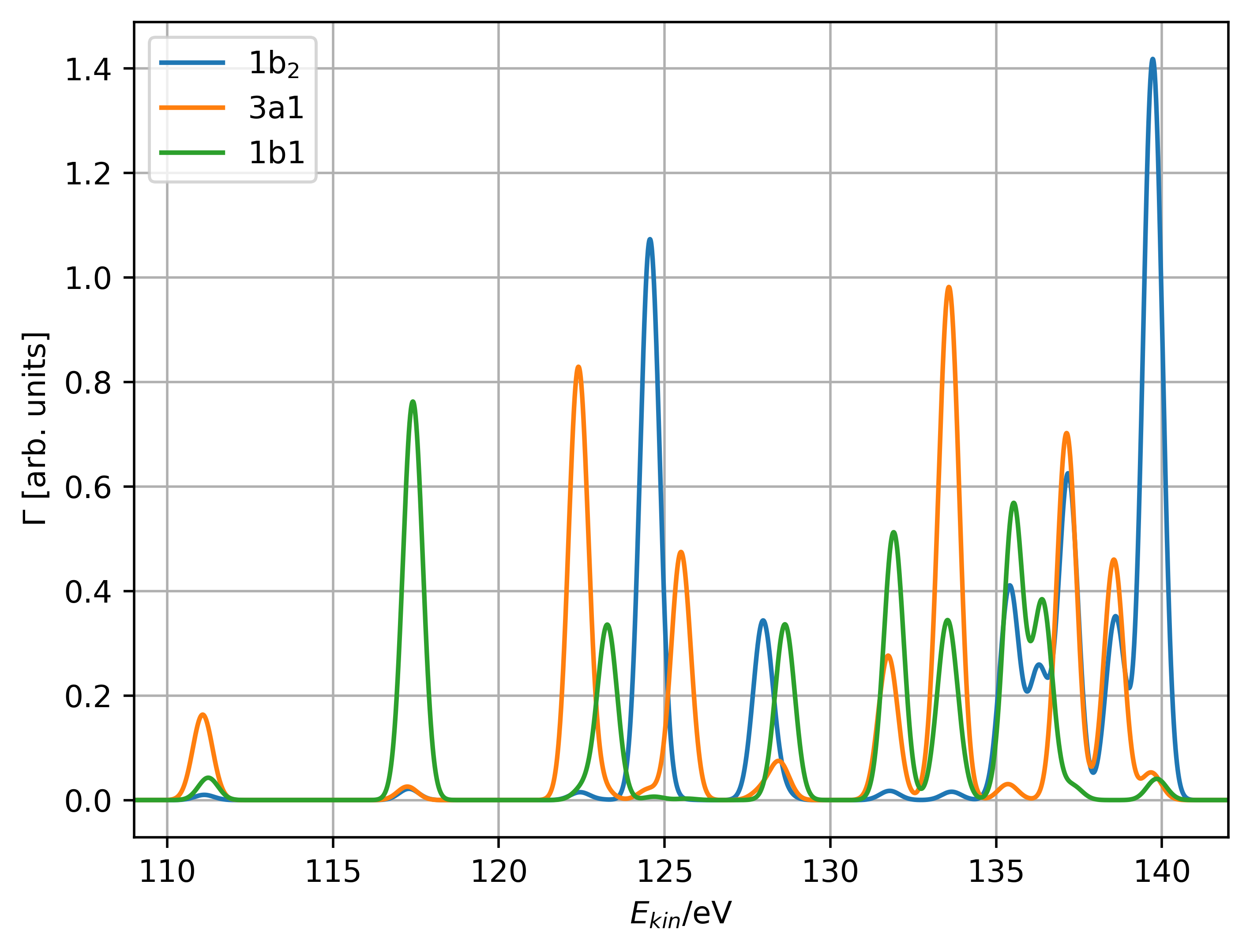}
\caption{$LMM$ Auger spectrum of hydrogen sulfide for the three 2p$^{-1}$ holes. Decay width
 in arbitrary units as a function of the kinetic energy of the emitted electron. 
 The partial widths were calculated with EOM-CCSD and 4(spdf) complex shells.
 The spectra are normalized to have the same relative sum of widths as in our calculation. Gaussian broadening with FWHM = 0.7 eV.}
\label{LMM H2S}
\end{figure}
The second part of the L-edge Auger spectrum corresponds to transitions arising from the 1b$_1^{-1}$, 3a$_1^{-1}$, and 1b$_2^{-1}$ core-holes. Experimentally, mostly the Auger transitions to 2b$_2^{-2}$ were investigated.\cite{poygin06,bueno03,svensson94} The reason is that these are the only transition that lead to a stable non-dissociating final state.\cite{poygin06} It has been suggested that this phenomenon occurs due to the 2b$_2$ orbital being the sole valence orbital oriented out of the molecular plane, rendering it nonbonding.\cite{poygin06} These authors also report the relative intensities of the various 
$\epsilon_Q \to $2b$_2^{-2}$ transitions, with Q = 1, 2, 3.
Poygin \textit{et al.} reported $0.35 \pm 0.05 : 0.04 \pm 0.01 :1$\cite{poygin06} whereas Bueno and coworkers obtained $0.42 \pm 0.03 : 0.09 \pm 0.03 :1$ experimentally and $0.42: 0.06:1$ theoretically.\cite{bueno03} Our calculated ratio after transforming our decay widths from the irrep basis to the 
$\epsilon$ states basis is $0.41: 0.06 :1$ which agrees excellently.

We can explain these ratios as follows. For Auger transitions of a non-s-type hole to be intense, one of the final holes should be parallel-oriented to the initial hole.\cite{gelmukhanov96} As the initial hole
$\epsilon_3^{-1}$
is mostly comprised of 1b$_2^{-1}$ and the 1b$_2$ orbital is, just as 2b$_2$, oriented out of the molecular plane, this respective transition is the most intense. The
$\epsilon_2$ orbital has almost no contribution from the 1b$_2$ orbital and the corresponding channel is therefore mostly suppressed.

This behavior can also be seen in all other transitions. We can see in our calculated partial widths, that for a channel to be open at least one of the final holes needs the same orientation as the initial hole. The spectra of the different holes therefore look extremely different as can be seen in figure \ref{LMM H2S}. Although the positions of the 16 peaks are also slightly different for the three holes, as they have slightly different ionization potentials, the reason for that is that different channels are open for the different holes. All three holes show seven significant peaks (relative partial width $>$ 0.1) each. For the 1b$_1^{-1}$ and 1b$_2^{-1}$ holes this is easily explained. There are just seven channels that include the 2b$_1$ and 2b$_2$ orbital, respectively. In the case of the 3a$_1^{-1}$ hole, the channels are open that include the 5a$_1$ orbital, which are also seven. The channels that include the 4a$_1$ orbital and not the 5a$_1$ orbital, however, are not open for this hole. The reason is that the 4a$_1$ orbital, despite having the same irrep as the 3a$_1^{-1}$ hole and being a valence orbital, still has significant characteristics of a s-orbital. It is thus not parallel oriented to the 3a$_1$ hole with p-character.
It should be noted that the presented spectra are not measurable because, as discussed, the three orbitals mix to form the three
$\epsilon$ states. Lastly, we want to point out that, as for the 2s$^{-1}$ hole, triplet channels contribute significantly to the total width. The triplet fraction to the sum of widths is 21\% (1b$_2^{-1}$), 28\% (3a$_1^{-1}$) and 30\% (1b$_1^{-1}$).

\section{Conclusions}
In conclusion, we calculated the total and partial Auger decay widths of the 1s$^{-1}$, 2s$^{-1}$ and 2p$^{-1}$ initial states of hydrogen sulfide and of the 2s$^{-1}$ state of argon. We used complex scaling of the Hamiltonian and complex basis functions to describe the emitted electron with $L^2$ integrable wave functions for argon and H$_2$S respectively. This work is an extension of this technique to the study of core orbitals from different shells.

For the 1s$^{-1}$ hole of H$_2$S, we found good agreement between the four used methods (CCSD and MP2 with decomposition and EOM-CCSD and EOM-MP2 with Auger channel projectors) and experimental and other theoretical values. We found that EOM-CCSD needs more CBFs to get a comparable sum of partial widths to the other methods. MP2 was used for the first time to calculate Auger widths with CBFs and gives very similar results, especially to CCSD (as both methods use the decomposition procedure), with much less computational costs. In the case of hydrogen sulfide, MP2 is about a factor 10 stronger than CCSD in a typical total and partial width calculation. It should be even more useful for larger molecules and complexes that cannot be tackled by CCSD at all.

Coster-Kronig transitions are present in the Auger decay from the 2s$^{-1}$ initial hole in H$_2$S and Ar. These are extremely strong due to the large spatial overlap of the 2s$^{-1}$ hole and the 2p electron that fills it, which leads to the very large decay widths of the 2s$^{-1}$ states. We saw that additional diffuse complex-scaled shells are needed to properly describe the slow emitted electrons in the Coster-Kronig transitions. In the case of argon we also performed CS calculations with good agreement to CBF, which verifies our approach, even though the exponents of the additional shells were not fully optimized.

In the Auger decay from the 2p$^{-1}$ states of hydrogen sulfide the total widths are one order of magnitude smaller than for the 1s$^{-1}$ hole and almost two orders smaller than for the 2s$^{-1}$ hole. The reason is that the 2p$^{-1}$ hole can only be filled by valence electrons which have a small spatial overlap with the core holes. We found that the three holes 1b$_1^{-1}$ 3a$_1^{-1}$ and 1b$_2^{-1}$ have very different spectra because different channels are open depending on the orientation of the initial hole. The governing rule is that one of the two final states needs to have the same spatial orientation as the initial hole.

We found that EOM-CCSD and EOM-MP2 are the only methods that work for the 2s$^{-1}$ and 2p$^{-1}$ holes. The latter, however, brings negligible computational benefit. The novel application of MP2 only gives trustworthy results for the holes, where CCSD converges.

In consensus with prior results\cite{jayadev23,matz22,matz23b,matz23a} we saw that triplet channels contribute very little to the total Auger width of the 1s$^{-1}$ state of H$_2$S ($5\%$ for all four methods). This is different for the other holes. We find that the 2s$^{-1}$ Auger width has a triplet contribution of 24\% for H$_2$S and 30\% for argon. For the three 2p holes of H$_2$S 1b$_2^{-1}$, 3a$_1^{-1}$ and 1b$_1^{-1}$ we find a triplet share of 21\%, 28\% and 30\%, respectively. The fact that singlet channels are still more important but triplet channels are not negligible in Auger decay from non 1s holes agrees with previous results.\cite{alkemper98}

As an outlook, we propose the implementation of spin-orbit coupling to the used code to achieve better agreement in the positions and heights of the Auger peaks with experiments. Furthermore, especially more recent experiments were dedicated to the shake structure and vibrational effects in Auger spectra\cite{puttner16,poygin06,bao08,pedersen23}, which are also not yet included in our description. Lastly, the potential of MP2 to describe the Auger decay in large molecules, that cannot be described by CCSD due to computational costs, should be explored in the future.

\section*{Acknowledgments}
T.-C.J. gratefully acknowledges funding from the European
Research Council (ERC) under the European Union’s Horizon
2020 research and innovation program (Grant Agreement No.
851766) and the KU Leuven internal funds (Grant No. C14/22/083).

\printbibliography

\end{document}